\documentclass{article}
\pdfoutput=1 
\usepackage{mathptmx}
\usepackage{helvet}
\usepackage{courier}
\usepackage[T1]{fontenc}
\usepackage[utf8]{inputenc}
\usepackage{amsmath}
\usepackage{amssymb}
\usepackage{graphicx}
\usepackage[authoryear]{natbib}
\usepackage[unicode=true,
 bookmarks=false,
 breaklinks=false,pdfborder={0 0 1},backref=false,colorlinks=false]
 {hyperref}
\hypersetup{pdftitle={Quantum measurement process with an ideal detector array},
 pdfauthor={Michael Zirpel},
 pdfsubject={Premeasurement with the result that exactly one detector is indicating a detection},
 pdfkeywords={measurement process, detector array, definite outcome, preferred base, wave function collapse}}

\makeatletter
\@ifundefined{date}{}{\date{}}
\makeatother

\begin{document}
\title{\noindent Quantum measurement process with an ideal detector array}
\author{Michael Zirpel, mz@mzirpel.de}
\maketitle
\begin{abstract}
Any observable with finite eigenvalue spectrum can be measured using
a multiport apparatus realizing an appropriate unitary transformation
and an array of detector instruments, where each detector operates
as an indicator of one possible value of the observable. The study
of this setup in the frame of von Neumann's quantum mechanical measurement
process has a remarkable result: already after the interaction of
the measured system with the detector array without collapse, exactly
one detector is indicating a detection. Each single detector indicates
either 0 or 1 detection, and no superposition can be attributed to
it. \textbf{} 
\end{abstract}

\section{Introduction}

Reck et al. \citeyearpar{ReckZeilinger1994} demonstrated that any
finite dimensional unitary transformation can be realized in the laboratory
using a lossless multiport apparatus constructed with beam splitters,
mirrors, and phase shifters. Furthermore, any finite dimensional observable
can be measured realizing an appropriate unitary transformation, attaching
a particle source to the input port and particle detectors to the
output ports. In this case, each detector operates as an indicator
of one possible value of the observable. 

Measurement instruments of this type are used in many modern versions
of quantum mechanical experiments. The authors list Einstein-Podolsky-Rosen
\citep{EPR1935,BohmAharonov1957} and Bell test \citep{Bell1964,CHSH1969}
experiments, as well as variants of Stern-Gerlach experiments \citep{GerlachStern1922},
and quantum cryptography devices \citep{Ekert1991}. A famous example
is Aspect's Bell test experiment \citep{Aspect1982} with photons,
which was repeated in many variants, even with neutrons \citep{Hasegawa2003}.
Several other experiments with neutrons (in line with their photon
counterparts) are described in \citep{Sponar2021}, with many additional
references.

In the following, such measurement instruments are considered theoretically,
with a focus on the detectors, which are treated quantum mechanically
in the frame of von Neumann's measurement process \citep{vNeumann1932,WheelerZurek1983}.
Any observable with finite eigenvalue spectrum can be measured with
an array of detector instruments using an appropriate measurement
interaction, described by a unitary transformation, which fulfills
the conditions for a repeatable ideal measurement. It is well-known,
that this interaction of the measured system and the measurement instrument
ends with an entangled state, which exhibits the ``measurement problem'':
how to get a definite outcome. Schrödinger's cat, waiting in a killing
machine triggered by the measurement outcome, is iconic for this situation:
is it dead, alive, or in a superposition of these states? For the
Copenhagen interpretation \citep{Heisenberg1958}, the measurement
must be completed by some final reduction or collapse \citep{BuschLahtiMittelstaedt1991}
to get a definite outcome. However, with the detector array the entangled
state already has some special properties, which are usually assumed
to arise with the collapse: \emph{exactly one detector is indicating
a detection}; \emph{each single detector indicates either 0 or 1 detection,
and no superposition can be attributed to it.}

\section{Von Neumann's measurement process}

Von Neumann \citeyearpar{vNeumann1932} described the measurement
process quantum mechanically as a short time interaction between the
measured system $S$ and a measurement instrument $M$, using Hilbert
space $\mathcal{H}_{S}$ and $\mathcal{H}_{M}$, respectively, as
state space, and the tensor product $\mathcal{H}_{S}\otimes\mathcal{H}_{M}$
for the composite system $SM$. The instrument has to display the
measurement outcome by entering one of the pairwise orthogonal pointer
states $\varphi_{1},\varphi_{2},...\in\mathcal{H}_{M}$, which correspond
to the eigenvalues $a_{1},a_{2},...\in\mathbb{R}$ of the measured
observable $A\in\mathcal{B}(\mathcal{H}_{S})$ (assumed to be bounded
and to have a pure, non-degenerate eigenvalue spectrum). The initial
state of the composite system $SM$ before the interaction and its
final state after that are connected by a unitary transformation $U_{SM}\in\mathcal{B}(\mathcal{H}_{S}\otimes\mathcal{H}_{M})$,
which must fulfill two conditions to constitute a repeatable ideal
measurement of the observable $A$: any eigenstate $\alpha_{k}$ of
$A$ stays unchanged, and the instrument must enter the corresponding
pointer state $\varphi_{k}$, indicating the measurement of the eigenvalue
$a_{k}$, i.e. 

\begin{equation}
U_{SM}\left(\alpha_{k}\otimes\varphi_{0}\right)=\alpha_{k}\otimes\varphi_{k}\,,\label{eq:vNeumannCondition}
\end{equation}
with $\varphi_{0}$ as the initial state of the instrument. As a consequence,
the interaction of the measurement instrument in initial state $\varphi_{0}$
with a system in state 
\[
\psi=\sum_{k}c_{k}\alpha_{k}\:,
\]
with $c_{k}\in\mathbb{C}$, transforms the composite system into the
entangled state 
\[
U_{SM}\left(\psi\otimes\varphi_{0}\right)=\sum_{k}c_{k}\alpha_{k}\otimes\varphi_{k}\,,
\]
with probability $p_{k}=\bigl|c_{k}\bigr|^{2}$ to observe the pointer
$\varphi_{k}$, according to Born's rule. This superposition of products
of eigenstates $\alpha_{k}$ and corresponding pointer states $\varphi_{k}$
exhibits the ``measurement problem'', how to get a definite outcome,
and is a reason for different interpretations of quantum mechanics
\citep{Mittelstaedt1998}. According to the Copenhagen interpretation,
however, this state terminates only the ``premeasurement'' \citep{BuschLahtiMittelstaedt1991}.
The measurement must be completed by some additional reduction or
collapse, giving a definite outcome in a random, irreversible transition
\[
\sum_{k}c_{k}\alpha_{k}\otimes\varphi_{k}\;\overset{p_{d}}{\curvearrowright}\;\alpha_{d}\otimes\varphi_{d}
\]
where $\varphi_{d}$ is the pointer state indicating the definite
outcome $a_{d}$, $\alpha_{d}$ the corresponding eigenstate of the
measured system and $p_{d}=\bigl|c_{d}\bigr|^{2}$ the transition
probability. Different authors associate this transition with different
circumstances, e.g. the recognition of the measurement outcome by
a conscious observer \citep{Wigner1961} or its registration by a
classical device \citep{Bohr1958}. Some non-Copenhagen interpretations
(e.g. the pilot wave interpretation \citep{Bohm1952,Bohm1952b}, the
many-worlds interpretation \citep{Everett1957}) deny such an additional
transition at all and assume that the final premeasurement state already
constitutes the end of the measurement.

The state $\alpha_{d}$ of the system after reduction is an eigenstate
of $A$, so a repetition of the measurement will give the same outcome
$a_{d}$. However, as Pauli \citeyearpar{Pauli1933} already noticed,
there are ideal measurements for which an immediate repetition gives
not the same result. The measurement scheme of the multiport apparatus
with detectors \citep{ReckZeilinger1994} is of this type. Von Neumann's
conditions (\ref{eq:vNeumannCondition}) must be relaxed to describe
such non-repeatable ideal measurements \citep{BuschLahtiMittelstaedt1991}:
an eigenstate $\alpha{}_{k}$ will be transformed into another state
$\alpha'_{k}\in\mathcal{H}_{S}$, while the instrument enters the
corresponding pointer state 
\begin{equation}
U_{SM}\left(\alpha{}_{k}\otimes\varphi_{0}\right)=\alpha'_{k}\otimes\varphi_{k}\,.\label{eq:Non-repeatable}
\end{equation}
For the multiport apparatus we can write $\alpha'_{k}=U\alpha{}_{k}$,
with the unitary transformation $U\in\mathcal{B}(\mathcal{H}_{S})$
determined by the apparatus. Here, these $\alpha'_{k}$ are not eigenstates
of $A$ but of the observable 
\begin{equation}
A'=UAU^{-1}\label{eq:ReckObservable}
\end{equation}
measured repeatably by the detectors. The collapse of the final premeasurement
state gives then
\begin{equation}
\sum_{k}c_{k}\alpha'_{k}\otimes\varphi_{k}\;\overset{p_{d}}{\curvearrowright}\;\alpha'_{d}\otimes\varphi_{d}\label{eq:NonRepCollapsedState}
\end{equation}
indicating the measurement outcome $a_{d}$, with the system in the
state $\alpha'_{d}$.

\section{Premeasurement with an ideal detector }

In the following, an ideal detector is considered as an instrument
for measuring an indicator observable (binary observable) which is
represented by a projection operator $E\in\mathcal{B}(\mathcal{H}_{S})$
with $E=E^{\dagger}=E^{2}$, eigenvalues $\sigma(A)=\{0,1\}$, and
the complement $\overline{E}=1-E$. A suitable detector instrument
$D$ has at least $2$ orthogonal pointer states $\varphi_{0},\varphi_{1}\in\mathcal{H}_{D}$
to display the measurement result. With Hilbert space $\mathcal{H}_{D}=\mathbb{C}^{2}$,
the detector instrument can be a spin-1/2-system or a qubit. 

The mathematical expression of von Neumann's conditions (\ref{eq:vNeumannCondition})
for the interaction $U_{SD}\in\mathcal{B}(\mathcal{H}_{S}\otimes\mathcal{H}_{D})$
must be generalized to include repeatable ideal measurements of degenerate
observables \citep{Lueders1951}. For an arbitrary vector $\psi\in\mathcal{H}_{S}$,
$E\psi$ and $\overline{E}\psi$ are (unnormalized) eigenvectors of
$E$ (or the zero vector) with eigenvalues $1$ or $0,$ respectively.
So, the conditions are

\begin{equation}
U_{SD}\left(E\psi\otimes\varphi_{0}\right)=E\psi\otimes\varphi_{1}\label{eq:vNeumannDectectorCondition}
\end{equation}
\[
U_{SD}\left(\overline{E}\psi\otimes\varphi_{0}\right)=\overline{E}\psi\otimes\varphi_{0}
\]
for all $\psi\in\mathcal{H}_{S}$.

\paragraph*{Example \label{exa:It-is-easy}1 }

Premeasurement with a detector. 

\noindent a) A vector of the Hilbert space $\mathcal{H}_{S}\otimes\mathbb{C}^{2}$
can be written as 2-component column vector (similar as a spinor).
In this notation, the matrix 
\[
V_{SD}=\left(\begin{array}{cc}
\overline{E} & E\\
E & \overline{E}
\end{array}\right)\in\mathcal{B}(\mathcal{H}_{S}\otimes\mathbb{C}^{2}),
\]
with $V_{SD}=V_{SD}^{-1}=V_{SD}^{\dagger}$, fulfills the conditions
for a repeatable ideal measurement (\ref{eq:vNeumannDectectorCondition}).\\
b) If the measured system $S$ is a spinless particle, with $\mathcal{H}_{S}=\mathcal{L}^{2}(\mathbb{R}^{3})$,
an indicator observable for position detection is the indicator function
$I_{V}:\mathbb{R}^{3}\rightarrow\{0,1\}$, with $I_{V}(r)=1\Leftrightarrow r\in V$,
of a Borel set $V\in B(\mathbb{R}^{3})$ representing the detection
volume (in position representation), with
\[
V_{SD}=\left(\begin{array}{cc}
1-I_{V} & I_{V}\\
I_{V} & 1-I_{V}
\end{array}\right)\in\mathcal{B}(\mathcal{L}^{2}(\mathbb{R}^{3})\otimes\mathbb{C}^{2})\,.
\]
c) If the measured system $S$ is a qubit \citep{NielsenChuang2000},
with $\mathcal{H}_{S}=\mathbb{C}^{2},$ a indicator observable for
a measurement in the computational base is 
\[
\tilde{E}=\bigl|1\bigr\rangle\bigl\langle1\bigr|=\left(\begin{array}{cc}
0 & 0\\
0 & 1
\end{array}\right),\overline{\tilde{E}}=\bigl|0\bigr\rangle\bigl\langle0\bigr|=\left(\begin{array}{cc}
1 & 0\\
0 & 0
\end{array}\right),
\]
and with its matrix representation one gets a $4\times4$ matrix for
the measurement transformation
\[
\tilde{V}_{SD}=\left(\begin{array}{cc|cc}
1 & 0 & 0 & 0\\
0 & \text{0} & 0 & 1\\
\hline 0 & 0 & 1 & 0\\
0 & 1 & 0 & 0
\end{array}\right),
\]
which is a representation of a CNOT circuit with 2 qubits (fig. 1).
\begin{center}
\begin{figure}[h]
\begin{centering}
\includegraphics[scale=0.5]{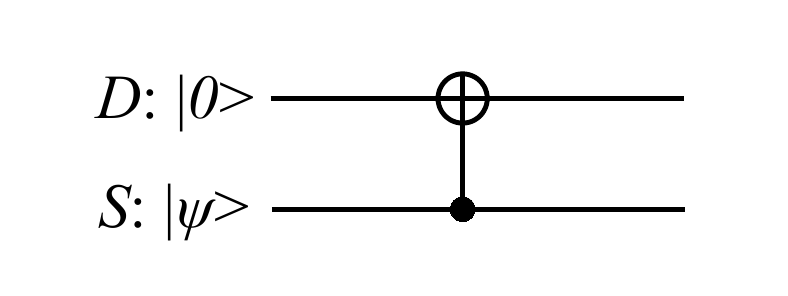}
\par\end{centering}
\caption{Premeasurement with CNOT circuit}
\end{figure}
\par\end{center}

\section{Premeasurement with an ideal detector array}

Any observable $A\in\mathcal{B}(\mathcal{H}_{S})$ with a finite eigenvalue
spectrum $\sigma(A)=\{a_{1},...,a_{n}\}\subset\mathbb{R}$ can be
written as a linear combination of projection operators $E_{k}\in\mathcal{B}(\mathcal{H}_{S})$
onto the eigenspaces
\[
A=\sum_{k=1}^{n}a_{k}E_{k}\,,
\]
with 
\begin{equation}
E_{j}E_{k}=E_{k}E_{j}=\delta_{j,k}1\label{eq:Eigenspaceorthogonality}
\end{equation}
for all $j,k\in I_{n}=\{1,...,n\}$, and 
\begin{equation}
\sum_{k=1}^{n}E_{k}=1\,.\label{eq:EigenUnitiy}
\end{equation}

The projection operators $E_{1},...,E_{n}$ are indicator observables
for the corresponding eigenvalues and can be measured using ideal
detector instruments $D_{1},...,D_{n}$, assuming a bijection
\begin{equation}
E_{k}\leftrightarrows D_{k}\,.\label{eq:BijectionDetectorProjector}
\end{equation}
An array of these $n$ detectors can be used as a measurement instrument
for the observable $A$. With pointer states $\varphi_{0}^{(k)}$,
$\varphi_{1}^{(k)}$ of detector $D_{k}$, the pointer states of this
instrument $M$ are
\[
\mathrm{\tilde{\varphi}}_{0}=\mathrm{\varphi_{0}^{(1)}}\otimes...\otimes\varphi_{0}^{(n)}\,,
\]
\[
\mathrm{\tilde{\varphi}}_{k}=\mathrm{\varphi_{0}^{(1)}}\otimes...\otimes\varphi_{1}^{(k)}\otimes\varphi_{0}^{(k+1)}\otimes...\otimes\varphi_{0}^{(n)}
\]
for $k\in I_{n}$, in Hilbert space $\mathcal{H}_{M}=\mathcal{H}_{D}^{(1)}\otimes...\otimes\mathcal{H}_{D}^{(n)}=\mathcal{H}_{D}^{\otimes n}$.
The measurement result is the index $k$ of the detector with the
pointer state $\varphi_{1}^{(k)}$, equivalent to the pointer state
$\mathrm{\tilde{\varphi}}_{k}$ of the whole array, indicating the
value $a_{k}$ of the observable $A$. Here, von Neumann's conditions
(\ref{eq:vNeumannCondition}) for an repeatable ideal measurement
of observable $A$ are 
\begin{equation}
U_{SM}(E_{k}\psi\otimes\tilde{\varphi}_{0})=E{}_{k}\psi\otimes\tilde{\varphi}_{k}\label{eq:Ideal Cond array}
\end{equation}
for all $\psi\in\mathcal{H}_{S}$ and $\ensuremath{k\in I_{n}}$.

In the rest of this paragraph, a construction for this unitary transformation
$U_{SM}\in\mathcal{B}(\mathcal{H}_{S}\otimes\mathcal{H}_{M})$ is
given, using the detector transformation $U_{SD}^{(k)}\in\mathcal{B}(\mathcal{H}_{S}\otimes\mathcal{H}_{D}^{(k)})$
from (\ref{eq:vNeumannDectectorCondition}) for each single detector
$D_{k}$. With (\ref{eq:Eigenspaceorthogonality}) we can write (\ref{eq:vNeumannDectectorCondition})
\begin{equation}
U_{SD}^{(k)}\left(E_{k}\psi\otimes\varphi_{0}^{(j)}\right)=E_{k}\psi\otimes\varphi_{\delta_{j,k}}^{(j)}\label{eq:DetectorArrayCondition}
\end{equation}
for all $j,k\in I_{n}$, and $U_{SM}$ can be constructed as product
\[
U_{SM}=U_{M}^{(1)}\cdot...\cdot U_{M}^{(n)}
\]
where 
\[
U_{M}^{(k)}=1^{(1)}\otimes...\otimes U_{SD}^{(k)}\otimes1^{(k+1)}...\otimes1^{(n)}
\]
is the representation of $U_{SD}^{(k)}$ in the product space $\mathcal{H}_{S}\otimes\mathcal{H}_{M}.$
It is obvious, that these operators commute 
\[
U_{M}^{(j)}U_{M}^{(k)}=1^{(1)}\otimes...U_{SD}^{(j)}\otimes...U_{SD}^{(k)}\otimes...\otimes1^{(n)}=U_{M}^{(k)}U_{M}^{(j)}
\]
for all $j,k\in I_{n}$. To demonstrate that $U_{SM}$ fulfills von
Neumann's conditions for the eigenstates $E_{k}\psi$ of $A$, we
multiply the initial state of the composite system 
\[
\varPhi_{0}=E_{k}\psi\otimes\mathrm{\varphi_{0}^{(1)}}\otimes...\otimes\varphi_{0}^{(n)}=E_{k}\psi\otimes\tilde{\varphi}_{0}
\]
for each $k\in I_{n}$ successively with $U_{M}^{(1)},...,U_{M}^{(n)}$
using (\ref{eq:DetectorArrayCondition})
\[
\varPhi_{1}=U_{M}^{(1)}\varPhi_{0}=E_{k}\psi\otimes\mathrm{\varphi_{0}^{(1)}}\otimes...\otimes\varphi_{0}^{(k)}\otimes...\otimes\varphi_{0}^{(n)}=E_{k}\psi\otimes\tilde{\varphi}_{0}
\]
\[
...
\]
\[
\varPhi_{k}=U_{M}^{(k)}\varPhi_{k-1}=E_{k}\psi\otimes\mathrm{\varphi_{0}^{(1)}}\otimes...\otimes\varphi_{1}^{(k)}\otimes\varphi_{0}^{(k+1)}\otimes...\otimes\varphi_{0}^{(n)}=E_{k}\psi\otimes\tilde{\varphi}_{k}
\]
\[
...
\]
\[
U_{SM}\varPhi_{0}=\varPhi_{n}=U_{M}^{(n)}\varPhi_{n-1}=E_{k}\psi\otimes\mathrm{\varphi_{0}^{(1)}}\otimes...\otimes\varphi_{1}^{(k)}\otimes\varphi_{0}^{(k+1)}\otimes...\otimes\varphi_{0}^{(n)}=E_{k}\psi\otimes\tilde{\varphi}_{k}\,,
\]
which gives (\ref{eq:Ideal Cond array}).

\paragraph*{Example 2 }

Premeasurement with a 2-detector array.

\noindent a) With the indicator observables $E_{1},E_{2}\in\mathcal{B}(\mathcal{H}_{S})$
the matrix representation of $V_{SD}$ from \nameref{exa:It-is-easy}
gives 
\[
V_{SM}=(V_{SD}^{(1)}\otimes1^{(2)})(1^{(1)}\otimes V_{SD}^{(2)})=\left(\begin{array}{cccc}
0 & E_{1} & E_{2} & 0\\
E_{1} & 0 & 0 & E_{2}\\
E_{2} & 0 & 0 & E_{1}\\
0 & E_{2} & E_{1} & 0
\end{array}\right),
\]
\[
V_{SM}\left(\begin{array}{c}
\psi\\
0\\
0\\
0
\end{array}\right)=\left(\begin{array}{c}
0\\
E_{1}\psi\\
E_{2}\psi\\
0
\end{array}\right),
\]
which fulfills (\ref{eq:Ideal Cond array}). \\
b) If the measured system $S$ is a particle, the Hilbert space can
be restricted to a finite box $B\in\mathcal{B}(\mathbb{R}^{3})$,
with $\mathcal{H}_{S}=\mathcal{L}^{2}(B)$. Then, the completeness
condition (\ref{eq:EigenUnitiy}) for the partition of the position
space $V_{1}\cup V_{2}=B$ can be fulfilled with two finite detection
volumes and the corresponding indicator functions $E_{1}=I_{V_{1}}$,
$E_{2}=I_{V_{2}}$ in position representation. Such partitions give
a simple model of a tracking chamber.\\
c) If the measured system is a qubit, the measurement interaction
for a measurement in the computational base with $\tilde{E}_{1}=\bigl|1\bigr\rangle\bigl\langle1\bigr|$
and $\tilde{E}_{2}=\bigl|0\bigr\rangle\bigl\langle0\bigr|$ can be
described with the resulting $8\times8$ matrix for $\tilde{V}_{SM}$
or, equivalently, with an $1$-to-$2$ decoder circuit (fig. 2).
\begin{center}
\begin{figure}[h]
\begin{centering}
\includegraphics[scale=0.5]{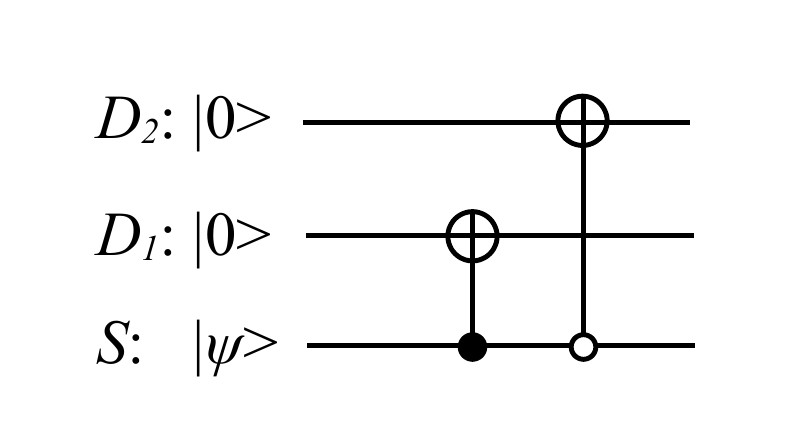}
\par\end{centering}
\caption{Premeasurement with 1-to-2 decoder circuit}
\end{figure}
\par\end{center}

\section{Properties of the final premeasurement state}

It is a consequence of (\ref{eq:EigenUnitiy}) and the conditions
(\ref{eq:Ideal Cond array}), that for an arbitrary initial state
$\psi\in\mathcal{H}_{S}$ of the measured system $S$ the final premeasurement
state of the composite system $SM$ is 
\[
U_{SM}\left(\psi\otimes\tilde{\varphi}_{0}\right)=U{}_{SM}\bigl(\bigl(\sum_{k=1}^{n}E_{k}\bigr)\psi\otimes\tilde{\varphi}_{0}\bigr)=\sum_{k=1}^{n}E{}_{k}\psi\otimes\tilde{\varphi}_{k}\:.
\]

To include non-repeatable ideal measurements with the multiport apparatus
(\ref{eq:Non-repeatable}), we discuss the more general final premeasurement
state 
\begin{equation}
\varPhi=\sum_{k=1}^{n}E'{}_{k}\psi\otimes\tilde{\varphi}_{k}\label{eq:FinalPremState}
\end{equation}
where $E'_{k}=U^{-1}E_{k}U$ is a projection operator for all $k\in I_{n}$,
defined by the inverse transformation of (\ref{eq:ReckObservable}). 

This entangled state of the system and the detector array has several
interesting properties.

\subsection{Exactly one detection}

An observable of the detector array is the number of detectors indicating
a detection $N\in\mathcal{B}(\mathcal{H}_{M})$. With the eigenvalue
spectrum $\sigma(N)=\{0,1,...,n\}$ it can be written as
\[
N=\sum_{k=0}^{n}kQ_{k}\,,
\]
with projection operators $Q_{k}\in\mathcal{B}(\mathcal{H}_{M})$
onto the closed subspaces spanned by the vectors
\[
V_{k}=\left\{ \varphi_{i_{1}}^{(1)}\otimes...\otimes\varphi_{i_{n}}^{(n)}\in\mathcal{H}_{M}\mid i_{1},...,i_{n}\in\{0,1\},i_{1}+...+i_{n}=k\right\} \,.
\]
It is obvious that $\mathrm{\tilde{\varphi}}_{0}\in V_{0}$ is an
eigenvector of $N$ with eigenvalue $0$ and $\mathrm{\tilde{\varphi}}_{1},...,\mathrm{\tilde{\varphi}}_{n}\in V_{1}$
are eigenvectors with eigenvalue $1$. Consequently, the final premeasurement
state $\mathrm{\varPhi}$ (\ref{eq:FinalPremState}) is an eigenstate
of $1_{S}\otimes N$ with eigenvalue 1
\[
\left(1_{S}\otimes N\right)\mathrm{\varPhi}=\sum_{k=1}^{n}E'_{k}\psi\otimes N\tilde{\varphi}_{k}=\sum_{k=1}^{n}E'_{k}\psi\otimes\tilde{\varphi}_{k}=\mathrm{\varPhi\,.}
\]
So, the observable $1_{S}\otimes N$ has in state $\mathrm{\varPhi}$
the dispersion free expectation value $1$. This means, exactly one
detector is indicating a detection. 

\subsection{Preferred pointer base states}

The biorthogonal Schmidt decomposition of an entangled state is in
general not unique \citep{Schlosshauer2004,Ekert1995}. Consequently,
the same entangled state can arise as a final premeasurement state
measuring incompatible observables. However, with the final premeasurement
state $\mathrm{\varPhi}$ (\ref{eq:FinalPremState}), only eigenstates
of the number operator with eigenvalue $1$ can be used as factors
for the detector array. This can be shown considering the alternative
Schmidt decomposition
\[
\varPhi=\sum_{j}d_{j}\beta_{j}\otimes\chi_{j}\,,
\]

\[
(1_{S}\otimes N)\varPhi=\varPhi\;\Rightarrow
\]

\[
\sum_{j}d_{j}\beta_{j}\otimes N\chi_{j}=\sum d_{j}\beta_{j}\otimes\chi_{j}\;\Rightarrow
\]
\[
\sum_{j}d_{j}\beta_{j}\otimes(N\chi_{j}-\chi_{j})=0\,.
\]
Because of the pairwise orthogonality, each summand must be zero.
Therefore, each $\chi_{j}$ has to be an eigenstate of $N$ with eigenvalue
$1$. 

There is no factorization of such an eigenstate with other single
detector states than the pointer states $\varphi_{0}$,$\varphi_{1}$.
To see this, let $\tilde{\varphi}\in\mathcal{H}_{D}^{\otimes n}$
be an eigenvector of the number operator with eigenvalue $1$ and
\[
\tilde{\varphi}=\chi\otimes\psi\,,
\]
a factorization with $\psi=a\varphi_{0}+b\varphi_{1}\in\mathcal{H}_{D}$.
Then, we can write
\[
\chi=c\chi_{0}+d\chi_{1}\,,
\]
with $N\left(\chi_{0}\otimes\varphi_{0}\right)=0$ and $N\left(\chi_{1}\otimes\varphi_{0}\right)=\chi_{1}\otimes\varphi_{0}$,
and 

\[
\chi\otimes\psi=ac\chi_{0}\otimes\varphi_{0}+ad\chi_{1}\otimes\varphi_{0}+bc\chi_{0}\otimes\varphi_{1}+bd\chi_{1}\otimes\varphi_{1}
\]

\[
N(\chi\otimes\psi)=0+ad\chi_{1}\otimes\varphi_{0}+bc\chi_{0}\otimes\varphi_{1}+2bd\chi_{1}\otimes\varphi_{1}\,.
\]
Non-trivial equality is only possible with $c=0$ and $b=0$ or with
$d=0$ and $a=0$. Since the factorization of the state 
\[
\varPhi=\sum_{j}c_{j}\beta_{j}\otimes\chi_{j}=\sum_{j}c_{j}\beta_{j}\otimes\tilde{\chi}_{j}\otimes\psi
\]
is a factorization of all $\chi_{j}$, it is only possible with $\psi\propto\varphi_{0}$
or $\psi\propto\varphi_{1}$.

For a single detector $D_{k}$ of the array, with $k\in I_{n}$, this
has the consequence that no superposition of the pointer states $\varphi_{0}^{(k)}$
and $\varphi_{1}^{(k)}$ can be attributed to the final premeasurement
state $\mathrm{\varPhi}$ (\ref{eq:FinalPremState}) as a factor of
a Schmidt decomposition. The pointer base states $\varphi_{0}^{(k)},\varphi_{1}^{(k)}$
are preferred \citep{Zurek2003} and the reduced state of each detector
$D_{k}$ is a unique mixture of $\varphi_{0}^{(k)}$ and $\varphi_{1}^{(k)}$\emph{.}

\subsection{Conditional expectations equal expectations after collapse}

Other studies \citep{LauraVanni2008} have shown, that for an ideal
repeatable measurement, fulfilling von Neumann's conditions, the final
premeasurement state gives the same conditional probabilities for
the measured system as the collapse of the wave function. This is
also the case with the collapse of the premeasurement state after
a non-repeatable ideal measurement (\ref{eq:NonRepCollapsedState}).

To see this for state $\mathrm{\varPhi}$ (\ref{eq:FinalPremState})
let $P_{k}\in\mathcal{H}_{M}$ be the indicator observable for detection
by detector $D_{k}$, i.e. the projection operator onto $\tilde{\varphi}_{k}$.
It commutes with all system observables. So, a common probability
space is available for all these observables and the conditional probability
of any event represented by the indicator observable $F\in\mathcal{B}(\mathcal{H}_{S})$
of the system, given $P_{k}\in\mathcal{B}(\mathcal{H}_{M})$, is well
defined for all $k\in I_{n}$

\[
p(F\mid P_{k})=\frac{p(F\wedge P_{k})}{p(P_{k})}=\frac{\bigl\langle\varPhi,F\otimes P_{k}\varPhi\bigr\rangle}{\bigl\langle\varPhi,1_{S}\otimes P_{k}\varPhi\bigr\rangle}=\frac{\bigl\langle\psi,E'_{k}FE'_{k}\psi\bigr\rangle}{\bigl\langle\psi,E'_{k}\psi\bigr\rangle}
\]
if $\bigl\langle\psi,E'_{k}\psi\bigr\rangle>0$. This is a consequence
of Born's rule. The state of the system after the complete measurement
with collapse, given the outcome $a_{k}$, is according to (\ref{eq:NonRepCollapsedState})
\[
\psi_{a_{k}}=\frac{1}{\sqrt{\bigl\langle\psi,E'_{k}\psi\bigr\rangle}}E'_{k}\psi\,.
\]
For this state the probability of $F$ is
\[
\bigl\langle\psi_{a_{k}},F\psi_{a_{k}}\bigr\rangle=\frac{\bigl\langle\psi,E'_{k}FE'_{k}\psi\bigr\rangle}{\bigl\langle\psi,E'_{k}\psi\bigr\rangle}=p(F\mid P_{k})
\]
and has the same value as the conditional probability of $F$, given
$P_{k}$, for the state $\varPhi$. The same is valid for the conditional
expectation value of any observable of the system$.$

Consequently, considering the Heisenberg picture, the isolated evolution
of the system $S$ after the measurement interaction, given $P_{k}$,
is the same as for the collapsed wave function after the measurement,
given the outcome $a_{k}$. 

\section{Conclusion}

The entangled state of the system and the detector array after the
ideal premeasurement of an observable has the remarkable property
that the number of detectors indicating a detection is exactly one.
This is a hint at a definite outcome of the measurement and resembles
the collapse, since the conditional expectations, given a detection,
are the same as for the collapsed state, given the corresponding outcome.
Furthermore, no superposition state can be attributed to a single
detector: the pointer base states and their mixtures are preferred. 

Using Schrödinger's cat illustration, one can say: if in a lossless
multiport experiment a particle has different exits for all possible
measurement outcomes, each one with an ideal detector connected to
a killing machine with its own cat, then exactly one cat will be dead,
and all others will be alive after a trial; there is no superposition
of dead and alive – even after a pure unitary evolution without collapse.

These properties depend on an ideal measurement process where the
detector array fulfills the completeness and orthogonality conditions
(\ref{eq:EigenUnitiy}), (\ref{eq:Eigenspaceorthogonality}) with
the bijection (\ref{eq:BijectionDetectorProjector}). Quantum computing
circuits (example 2c) demonstrate the possibility of such a setting.
In case of position detectors (example 2b), exactly positioned detection
volumes without gap and overlap, filling the available space, are
required, which seams only possible in a macroscopic quasi-classical
limit, in line with Bohr's Copenhagen \citep{Bohr1958} interpretation. 

Solely, these properties cannot constitute irreversible facts. That,
however, could be accomplished through the decoherence of the final
premeasurement state \citep{Schlosshauer2004}.

The results of our study may support interpretations of quantum mechanics
without collapse. However, if some of the latter's observable consequences
are deducible from other rules, this will not mean a contradiction
to the Copenhagen interpretation. Even, if the collapse postulate
was rendered unnecessary, it could persist as useful rule of thumb. 

\bibliographystyle{plainnat}

\end{document}